# Cluster Generation Under Pulsed Laser Ablation Of Compound Semiconductors

Alexander V. Bulgakov[a], Anton B. Evtushenko[a], Yuri G. Shukhov[a], Igor Ozerov[b] and Wladimir Marine[b]

[a]*Institute of Thermophysics SB RAS, Lavrentyev Ave. 1, 630090 Novosibirsk, Russia*
[b]*Université de la Méditerranée, CINaM, UPR CNRS 3118, 13288 Marseille, France*

**Abstract.** A comparative experimental study of pulsed laser ablation in vacuum of two binary semiconductors, zinc oxide and indium phosphide, has been performed using IR- and visible laser pulses with particular attention to cluster generation. Neutral and cationic $Zn_nO_m$ and $In_nP_m$ particles of various stoichiometry have been produced and investigated by time-of-flight mass spectrometry. At ZnO ablation, large cationic (n>9) and all neutral clusters are mainly stoichiometric in the ablation plume. In contrast, indium phosphide clusters are strongly indium-rich with $In_4P$ being a magic cluster. Analysis of the plume composition upon laser exposure has revealed congruent vaporization of ZnO and a disproportionate loss of phosphorus by the irradiated InP surface. Plume expansion conditions under ZnO ablation are shown to be favourable for stoichiometric cluster formation. A delayed vaporization of phosphorus under InP ablation has been observed that results in generation of off-stoichiometric clusters.

## INTRODUCTION

Efforts to develop methods for controllable synthesis of nanoclusters and nanostructured thin films of multicomponent materials are motivated by both fundamental science and the prospects for applications. In particular, nanostructured materials of II-IV and III-V compound semiconductors like ZnO, CdTe, InP have great potential in various technological applications such as solar cells, chemical sensors, photocatalysis, and nanoscale optoelectronic devices [1-4]. All these applications require a very high quality of the materials with respect to purity, defect concentration, nanoscale uniformity. However, producing multicomponent nanoclusters with desirable properties is still challenging. As compared to single-component materials, there is an additional key parameter in this case, cluster composition, which strongly affects cluster properties and makes the fabrication process more difficult to control and optimize.

Pulsed laser ablation (PLA) has proven to be an efficient and flexible method for synthesis of nanostructures of various materials including simple and compound semiconductors. Although the physical processes involved in laser ablation are very complex, important progress has been achieved in recent years in PLA synthesis of nanoclusters of simple semiconductors with controlled size, shape and optical properties [5,6] and in characterization of the cluster formation process [7-9]. However, mechanisms of compound cluster generation under PLA conditions are still poorly understood. Despite of numerous reports on PLA-produced nanostructures and nanostructured films of binary semiconductors (see, e.g., [2,10-12]), little is known of the origin of the nanoclusters which can be either directly ejected from the target or formed in the ablation plume via a condensation process during the plume expansion [3,9]. The existing data on the composition of compound clusters in the PLA plume are rather contradictory. Thus, highly off-stoichiometric (zinc-rich) clusters were observed in Ref. [13] under PLA of ZnO in vacuum while other works report on production of stoichiometric [14] and even oxygen-rich [15] clusters under similar ablation conditions. Indium phosphide clusters produced by PLA in a carrier gas were found to be highly indium-rich [16] and essentially stoichiometric [17]. The cluster size and stoichiometry depend on various PLA parameters such as laser intensity and wavelength as well as ambient gas environment [10,18-20]. Moreover, the composition the irradiated sample can vary from pulse to pulse due to non-congruent vaporization resulting in enrichment of a surface layer with a less volatile component [21,22]. The existing data on thermal evaporation of compound semiconductors demonstrate that there are two different ways of their decomposition. Some two-component materials, e.g., ZnO, vaporize congruently over a wide temperature range [23] whereas other semiconductors, for instance, InP and GaAs, exhibit a disproportionate loss of high-volatile components when temperature exceeds a critical point [24]. Detailed mechanisms of the non-congruent vaporization are still not clear even for the relatively simple situation of thermal decomposition [25].

In this work, we have performed a comparative mass-spectrometric study on generation of clusters of two binary semiconductors, zinc oxide and indium phosphide, under PLA in vacuum with IR (1064 and 800 nm) and visible (532 nm) laser pulses. We focus on investigation of small clusters, both charged and neutral, which are formed in early ablation stages and play a role of nucleation centres for further nanocluster growth [7]. Particular attention is paid to the expansion dynamics of different plume constituents and to the plume composition change during multi-pulse irradiation.

## EXPERIMENTAL

The apparatus used for laser ablation and cluster production and detection was described earlier [9,26-28]. The target (a sintered polycrystalline ZnO pellet or a (100) InP wafer) was placed in a rotating holder in a vacuum chamber (base pressure $10^{-5}$ Pa) and irradiated by a laser pulse. Three different laser systems operating at 1064 nm (9 ns pulse, Nd:YAG laser), 532 nm (7 ns, 2nd harmonic of Nd:YAG laser), and 800 nm (100 fs, Ti:sapphire laser) were used for ablation. The laser fluence at the target was varied in the range 0.3 – 8 J/cm$^2$ for ZnO ablation and 0.05-1 J/cm$^2$ for InP ablation. The relative abundances of neutral and cationic particles in the PLA plume were analyzed by a reflectron time-of-flight mass spectrometer (RETOF MS). The apparatus allows studying neutral and charged species independently. The neutrals were pulse-ionized in the ion source of the MS either by electron impact (1 μs, 90 eV) or by 193-nm photons (5-ns laser pulse of an ArF laser). When investigating neutrals, two ion-repelling plates [28] were used to prevent the proper plume ions from reaching the ion source of the RETOF MS.

The plume expanded under field-free conditions over a distance of 6 cm (at ns-ablation) or 11 cm (at fs-ablation) towards a repeller grid where the plume ions (or pulse-ionized neutrals) were sampled by pulsing the grid at a time delay $t$ after the laser pulse. Variation of the time delay allowed the characterization of the plume expansion dynamics and temporal evolution of the plume composition. When studying clusters, a mass filter in front of the MS detector was used in order to suppress the huge atomic ($Zn^+$ or $In^+$) peaks and thus to avoid detector saturation. A voltage pulse applied to the filter blocked the detector for a time interval when the main plume ions arrived. High resolution of the RETOF MS (~1000) allowed us to distinguish between $Zn_nO_m$ and $Zn_{n+1}O_{m-4}$ according to their isotope distributions. Every mass spectrum was averaged over 200 laser shots unless otherwise specified.

## RESULTS AND DISCUSSION

We carried out three kinds of experiments: (a) studies of relative abundances of neutral and charged plume particles as a function of laser fluence with a focus on cluster distributions; (b) mass spectrometric investigations of the expansion dynamics of the major plume constituents, and (c) measurements of the yields of various plume particles as a function of the number of consecutive laser pulses applied at a fixed fluence.

### Clusters Produced by Laser Ablation of Zinc Oxide

For all studied conditions, neutral Zn atom and $O_2$ molecule were the dominant plume particles and positively charged $Zn^+$ was the dominant ion though the plume ionization degree was not exceeded ca. 1% throughout the laser fluence ranges studied. Other species, less abundant but always presented in the plume, were neutral O atom, and $O^+$, $O_2^+$, and $Zn_2^+$ ions. At relatively narrow ranges of laser fluences, rich spectra of neutral and charged zinc oxide clusters were observed for both IR- and visible-laser ablation. The optimum laser fluence for cluster production under vacuum conditions is found to increase with wavelength being in the range approx. 1-3 J/cm$^2$ for 532 nm and 2-5 J/cm$^2$ for 1064 nm.

Figure 1 shows a typical mass spectrum of cationic clusters observed at 1064-nm laser ablation when cluster ions are especially abundant. Under optimal ablation conditions, fairly large $Zn_nO_m^+$ clusters (up to n = 35) are registered and the cluster fraction can rich up to 10% of the total number of charged particles in the plume. The clusters are produced in abundance not only with a virgin target surface, as was in the case of UV-ablation [20], but also after etching the sample with several thousand pulses (though we avoided considerable cratering at the surface). The measurements with varying time delay $t$ show that the observed clusters have near equal expansion velocities ~ 1.5 km/s in spite of rather large difference in their mass. This implies a gas-phase aggregation mechanism for their formation rather that direct ejection [9,28]. The distribution has no obvious preferential (magic) numbers but the cluster composition strongly depends on the cluster size. The smallest species (n = 2, 3) are essentially substoichiometric (Zn-rich). The intermediate-size particles (n = 4-9) occur mainly as a twin peak, the first peak corresponding to $Zn_nO_{n-1}^+$ clusters with one missing oxygen atom while another peak is for stoichiometric species. The twin clusters seem to follow a rule that the first peak is higher at n = odd, while the second peak dominates at n = even. Large clusters (n > 9) are preferentially stoichiometric. Clusters with a distribution similar to that shown in Fig. 1 (though less abundant) were produced in this work by 532-nm ablation with laser fluence around 2 J/cm$^2$ and were observed for 337-nm ablation of ZnO [14]. On the other hand, a completely different (much poorer) cluster distribution with preferential sizes was reported for 1064-nm ablation of ZnO in vacuum [15]. This is probably due to the high laser fluence (7.7 J/cm$^2$) used in [15] which is beyond the optimum fluence range found in this work.

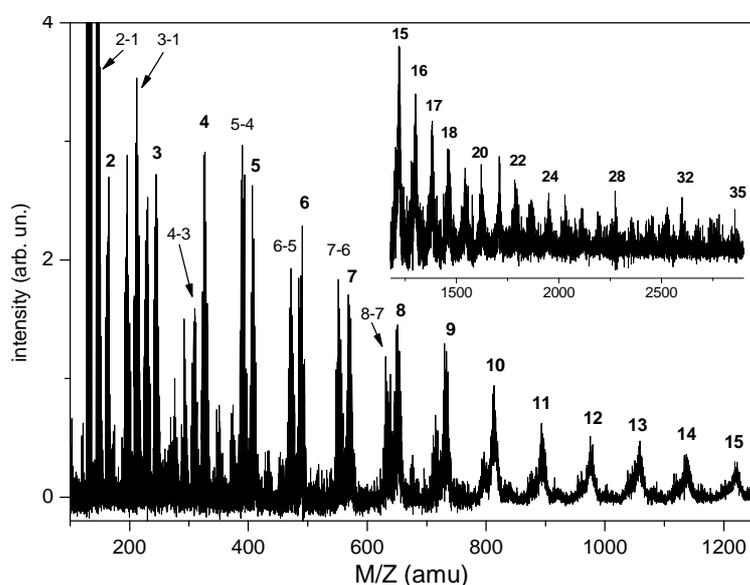

**FIGURE 1.** Mass spectrum of cationic clusters produced by ZnO ablation with 1064-nm laser pulses at 3 J/cm$^2$. The pairs of numbers (n-m) above the peaks correspond to the numbers of Zn and O atoms in $Zn_nO_m^+$ clusters. The single numbers above the peaks stand for *n* in stoichiometric $(ZnO)_n^+$ clusters.

Figure 2 shows the mass spectrum of neutral particles obtained with electron impact ionization and at the same "optimum" ablation conditions of ZnO as for cationic cluster generation (Fig. 1). The neutral Zn, $O_2$ and O species are totally dominant in the plume but, using the mass filter, we were able to detect neutral $Zn_nO_m$ clusters up to n = 7. The clusters are essentially stoichiometric at n > 3 while substoichiometric $Zn_4O_3$ and $Zn_5O_4$ species are registered. The most intriguing feature of the spectrum shown in Fig. 2 is the strong dominance of the $Zn_4O_4$ tetramer over the neighbors. It is rather surprising since recent DFT calculations of zinc oxide cluster structures [29] do not predict special stability for the tetramer. We suggest that the appearance of the $Zn_4O_4$ specie as a magic cluster in the mass spectrum is due to fragmentation of larger plume clusters upon electron impact. An argument for the fragmentation origin is that the $Zn_4O_4$ unit is one of the main building blocks of $Zn_nO_n$ clusters [29]. We performed a mass spectrometric study of neutral clusters under the same ablation conditions using the photo ionization technique with 6.4-eV photons (ArF laser) instead of electron impact. However this resulted in even stronger decomposition and we were not able to detect particles larger than dimers. Such a strong photofragmentation of zinc oxide clusters indicates that the ionization potential (IP) of large clusters (which presumably exist in the plume) is still higher than 6.4 eV and thus the probability of cluster decay due to absorption of two photons on the ns timescale is very high. Indeed, according to [30], IP of $Zn_nO_n$ clusters is about 7.7 eV in the n = 10-16 range). It would be interesting to investigate neutral particles in the ZnO ablation plume under single-photon ionization conditions, e.g. with $F_2$ laser (photon energy 7.9 eV). Preparation of such experiments is under progress.

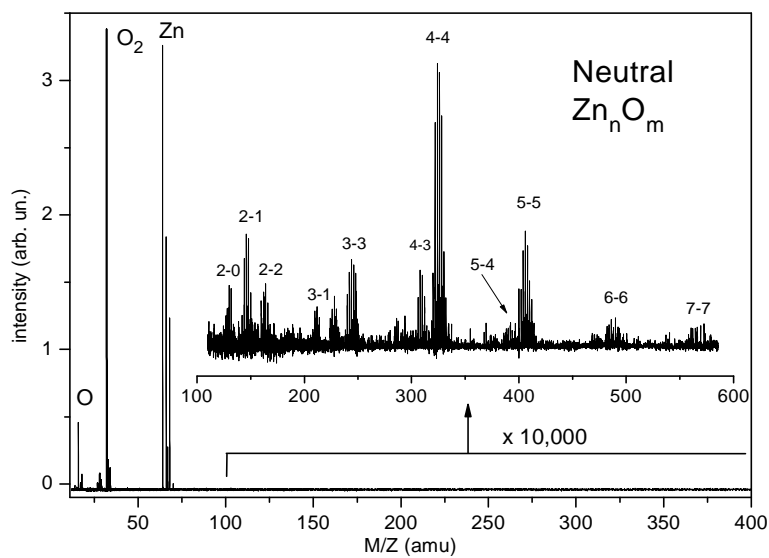

**FIGURE 2.** Mass spectrum of neutral particles produced by 1064-nm laser ablation of ZnO and pulse-ionized by electron impact at 90 eV. Ablation conditions are same as for Fig. 1.

## Clusters Produced by Laser Ablation of Indium Phosphide

Similar to ZnO ablation, three particles are dominant in the PLA-produced plume above the InP target throughout the irradiation conditions studied, namely, neutral In atom and $P_2$ molecule, and $In^+$ ion. Other species presenting in the plume at lower abundances are neutral $P_3$ and $P_4$ molecules, P atoms (the latter is observed only at fairly high fluences), and $In_2^+$ ions. Due to a low ionization potential of indium (5.8 eV), the fraction of charged particles in the plume is considerably higher than that under ZnO ablation and the plume ionization degree can reach up to 10% for the highest laser fluences investigated. Correspondingly, the thresholds for ablation and plasma formation are relatively low being ca. 150 mJ/cm$^2$ for 532-nm pulses, ca. 300 mJ/cm$^2$ for 1064-nm pulses, and ca. 100 mJ/cm$^2$ for 800-nm fs-laser pulses.

Again similar to ZnO ablation, at narrow fluence ranges of around 200-400 mJ/cm$^2$ for 532-nm and 400-700 mJ/cm$^2$ for 1064 nm, compound $In_nP_m$ clusters are observed in the plume. At fs-laser ablation, indium phosphide clusters are particularly abundant in the plume and registered in a wide fluence range starting almost from the ablation threshold (~100 mJ/cm$^2$). However, in contrast to the ZnO case, the clusters are now highly off-stoichiometric and their generation depends strongly on the laser wavelength. Under visible-laser ablation, charged clusters have been produced only with a relatively fresh target exposed to less than ~ 20 laser shots. By measuring the plume composition after every consecutive laser shot applied to the same spot we have found that the most pronounced cluster peaks are observed after 3rd and 4th laser pulses and drop drastically in intensity upon further laser exposure. The first two pulses serve to prepare the surface for cluster emission and to remove trace contaminations. Figure 3 shows a mass spectrum of cationic particle observed at 532-nm laser ablation of InP after 4th pulse. In contrast to the rather smooth distribution of zinc oxide cluster ions (Fig. 1), only four magic $In_nP_m^+$ clusters with n = 4-6 and m = 1-3 are present in the plume. Interesting that namely these clusters were observed in a mass spectrometric study of neutral products of 532-nm laser ablation of InP in a carrier gas with an ArF ionizing laser [16]. Therewith no neutral $In_nP_m$ clusters were observed in the present work under 532-nm PLA of InP in vacuum.

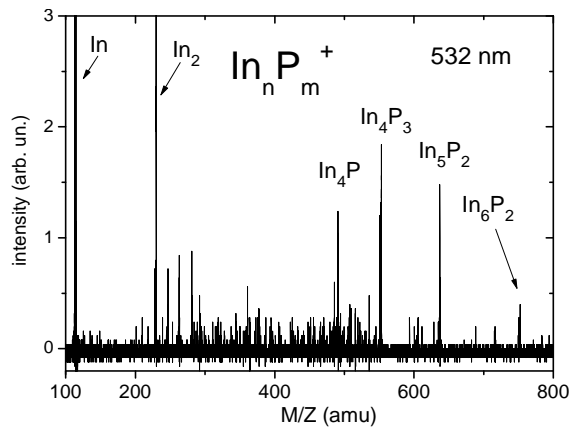

**FIGURE 3.** Mass spectrum of cationic clusters produced by InP ablation with 532-nm laser pulses at 280 mJ/cm$^2$. The spectrum was collected after 4th laser pulse applied to a virgin InP surface.

The situation is completely different under IR-laser ablation of InP (with both ns and fs pulses) resulting in rich spectra of neutral and charged clusters. Indium oxide clusters are now observed not only with a fresh target surface but also after etching the surface with several thousand pulses. Figure 4 shows the distribution of cationic $In_nP_m^+$ clusters produced under 800-nm fs-laser ablation when they are especially abundant. All observed clusters are essentially indium-rich and contain no more than 3 phosphorus atoms even for largest clusters. Depending on the In/P ratio the clusters may be gathered into two major series, $In_nP$ and $In_nP_2$ with (n+m) = odd (except for $In_6P_2$). The most abundant (magic) cationic clusters are the same as observed at 532-nm ablation (see Fig. 3), $In_4P^+$, $In_4P_3^+$, $In_5P_2^+$, and $In_6P_2^+$. Similar to zinc oxide clusters, all observed indium $In_nP_m^+$ cations have near equal expansion velocity ~ 2 km/s that again suggests a gas-phase aggregation mechanism for their formation. Ablation of InP by 1064-nm laser pulses results in generation of ionized clusters with a similar distribution but considerably less abundant. This is not surprising since fs ablating pulses are generally more favorable for gas-phase cluster production as compared to ns-laser pulses due to absence of plasma absorption effects and a more forward-directed plume resulting in more particle collisions during plume expansion [9,31].

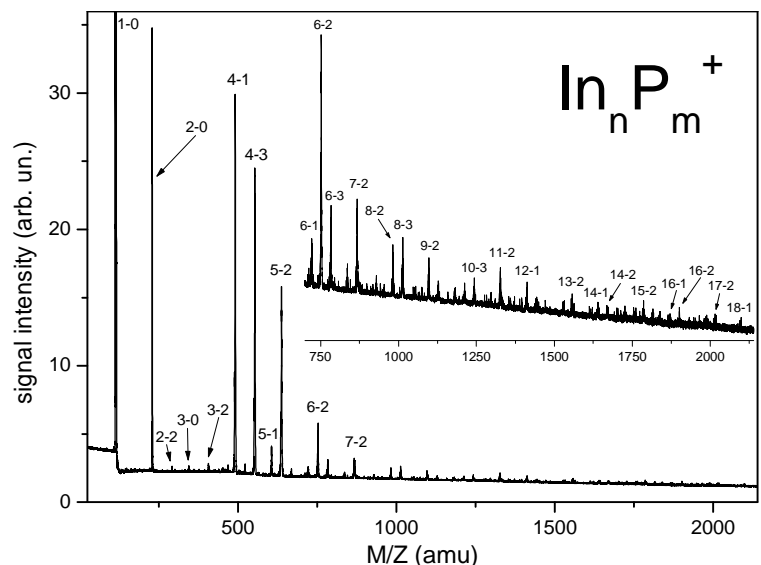

**FIGURE 4.** Mass spectrum of cationic $In_nP_m^+$ clusters produced by InP ablation with 800-nm fs-laser pulses at 260 mJ/cm$^2$. The numbers (n-m) above the peaks stand for the *n* and *m* values.

Figure 5 shows a mass spectrum of neutral particles obtained with photoionization (ArF laser) at IR-laser ablation of InP. The ionization potentials of $In_nP_m$ clusters appear to be below 6.4 eV (at least for $n > 3$) [16] and thus ionization of the clusters is a single-photon process without their considerable fragmentation. So the observed cluster distribution is expected to correspond to the true distribution of neutrals. Again the neutral clusters are strongly off-stoichiometric. By comparing the neutral cluster compositions with those of cationic clusters (Fig. 4) one can see that, for the same *n* value, most of ions have an additional P atom. However, the two magic clusters, $In_4P$ and $In_5P_2$ are dominant among both neutrals and ions. The enrichment of the observed $In_nP_m$ clusters with indium is explained by peculiarities of laser-induced vaporization and plume expulsion under PLA of InP as discussed below.

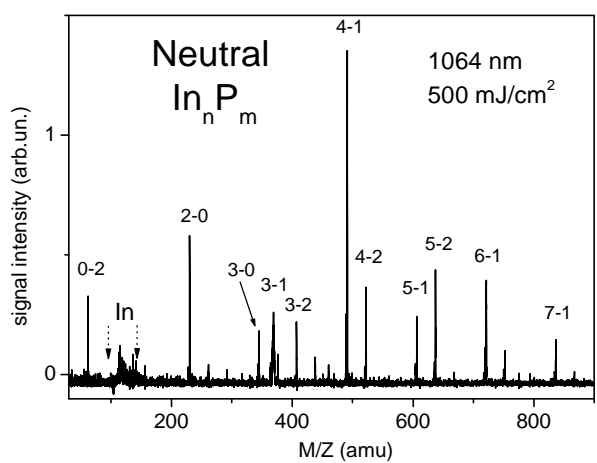

**FIGURE 5.** Mass spectrum of neutral clusters produced by InP ablation with 1064-nm laser pulses at 500 mJ/cm$^2$. The clusters were ionized by unfocused 193-nm laser pulses at ~20 mJ/cm$^2$. The $In^+$ ions were blocked from reaching the detector using a filter (the blocked interval is shown by the arrows).

# Shot Number Dependence of the PLA Plume Composition

To follow the pulse-to-pulse evolution of the vaporization process, we measured the yields of the main ablation products starting from virgin surfaces. The experiments were performed at fixed laser fluences and at time delays $t$ corresponding to the highest signals of neutral particles (at maxima of the TOF distributions). Figure 6a shows the results for Zn atoms and $O_2$ molecules produced by 1064-nm laser ablation of ZnO at a fluence of 3 $J/cm^2$ corresponding to the optimum conditions of cluster generation. The particle yields decrease first upon laser exposure, reach a minimum at ~30 shots, then increase again reaching a quasistationary level after ~1000 shots. Similar dependences were obtained at lower laser fluences (when, however, more laser shots are needed to reach a stationary value) and with 532-nm ablating pulses. Such a nonmonotonic dependence can be attributed to a complicated modification of the ZnO surface during irradiation resulted in variations of its optical properties (e.g. due to laser-induced generation of oxygen vacancies [10]). More important in the context of cluster generation is that the ratio of the zinc and oxygen yields remains constant (~ 1 within the experimental error). This means that vaporization of the ZnO target under these conditions occurs congruently, much like as thermal vaporization [23]. In its turn this implies a thermal nature of ZnO ablation with IR- and visible laser pulses. Note that this does not appear to be the case for UV-laser pulses when the surface is enriched with zinc upon irradiation [32] and Zn-rich clusters are observed [20].

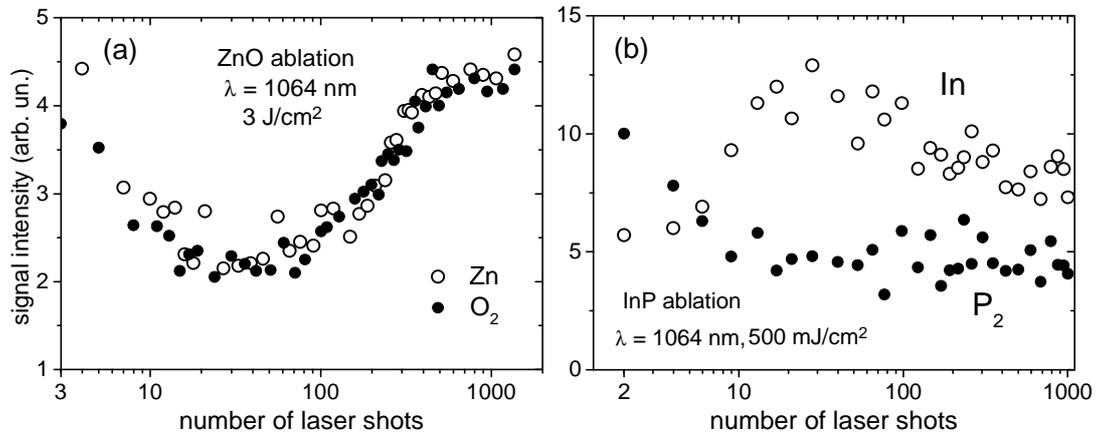

**FIGURE 6.** Yields of the main plume particles under 1064-nm laser ablation as a function of the number of laser shots. (a) is for neutral Zn atoms and $O_2$ molecules produced at ZnO ablation at a laser fluence of 3 $J/cm^2$; (b) is for neutral In atoms and $P_2$ molecules produced at InP ablation at a laser fluence of 500 $mJ/cm^2$. The yields for all particles are obtained at a time delay $t = 30$ μs.

A different evolution of the plume composition is observed at InP ablation. Figure 6b illustrates the yields of the main plume particles in this case, neutral In and $P_2$, as a function of the laser shot number for 1064-nm pulses. At a virgin InP surface, the $P_2$ signal is considerably stronger than that of In atoms (the $P_2$/In ratio is about 2 for this conditions and even higher for lower fluences). Upon irradiation, the phosphorus yield drops quickly while that of indium increases and after several hundred of shots both signals reach quasistationary levels with the $P_2$/In ratio being about 0.5. Similar shot-number dependences at InP ablation were obtained in this work with 532-nm pulses and in [33] at 337-nm ablation of InP where such a behavior was explained by the preferential emission of more volatile phosphorus that makes the irradiated surface In-rich. Indeed, the disproportionate loss of phosphorus and alteration of the surface stoichiometry were previously observed for thermal heating [24] and visible-laser irradiation [34] of InP samples. We note, however, that the observed quasistationary value of the $P_2$/In ratio after a fairly large number of laser shots (Fig. 6b) suggests equal overall yields of phosphorus and indium under these conditions even with the In-reach surface. This can be possible if phosphorus and indium are released from the surface at different timescales that is the case as we will see below.

## Plume Expansion Dynamics

To further elucidate the laser-induced vaporization and cluster formation process, the temporal evolution of the dominant plume particles was examined by varying the time delay $t$. Figure 7a shows the time-of-flight (TOF) distributions of neutral Zn and O atoms and $O_2$ molecules obtained under "optimum" conditions for ZnO ablation (1064 nm, 3 J/cm$^2$) resulting in the spectra shown in Figs. 1 and 2. As seen, the main plume components have essentially the same expansion velocity (around 2 km/s) and expand jointly. This implies a hydrodynamic regime of expansion during a prolonged time [26] that is favorable for stoichiometric cluster formation. Gas-phase collisions of different plume particles under these conditions are equiprobable and thus clusters have a high probability to be self-organized into most stable structures which are apparently stoichiometric in this case.

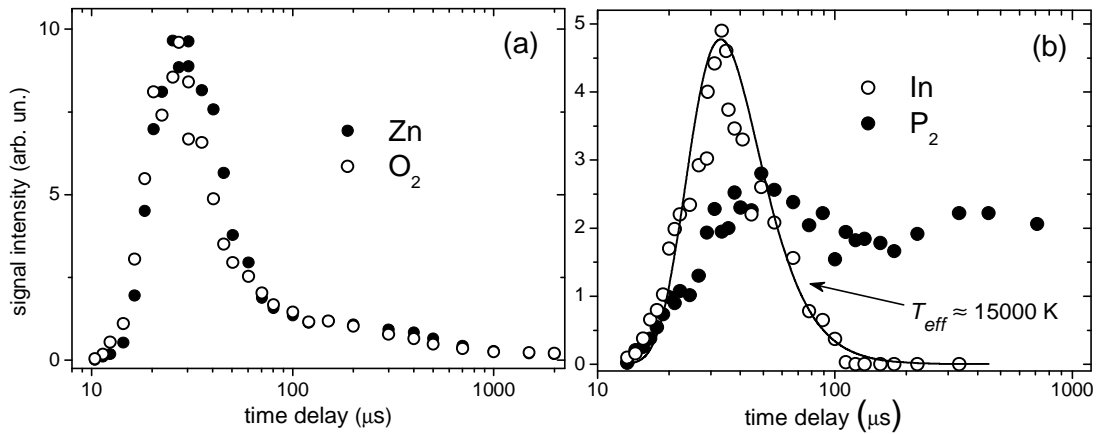

**FIGURE 6.** Time-of-flight distributions of the main plume particles under 1064-nm laser ablation. (a) is for neutral Zn and $O_2$ produced at ZnO ablation at a laser fluence of 3 J/cm$^2$; (b) is for neutral In and $P_2$ produced at InP ablation at a laser fluence of 500 mJ/cm$^2$. The solid line in (b) corresponds to a Maxwell-Boltzmann fit of the indium distribution with an effective plume temperature of 15000 K.

In contrast to the zinc oxide PLA plume, the main products of InP ablation products exhibit very different expansion behaviors. Figure 7b shows the TOF distributions of In atoms and $P_2$ molecules obtained with 1064-nm laser pulses at a fluence corresponding to maximum cluster yields (see Fig. 5). The In distribution is quite similar to those for particles produced under ZnO ablation (see Fig. 7a): it is single-peaked and rather narrow, is maximized at ca. the same arrival time of ~ 30 μs

(corresponding velocity ~2 km/s) and can be fitted with the Maxwell-Boltzmann distribution with a reasonable effective plume temperature of about 15000 K (solid line in Fig. 7b). The lighter $P_2$ molecules arrive therewith to the detector at considerably later times and their TOF distribution has a clear delayed tail corresponding to a fraction of very slow molecules. This delayed fraction of $P_2$ particles is found to increase with laser fluence and become dominant at above ~700 mJ/cm$^2$. Similar TOF distributions were obtained at 532-nm ablation of InP. Such peculiarities of the plume expansions under multi-pulse ablation of InP can be explained using above considerations of the overall stoichiometric vaporization from the irradiated In-rich surface. Indeed, indium is released via the *normal* vaporization mechanism [35], mainly during the laser pulse action when the surface temperature is maximal. The heating of the sample stimulates diffusion processes within an off-stoichiometric surface layer [22] resulting in a partial compensation of the phosphorus lack at the surface. Then the low-volatile phosphorus evaporates for a quite long time during cooling down the sample, leaving thus the surface again essentially indium-reach. Such a diffusion-limited vaporization produces rather slow particles and manifests itself by the delayed tail in the TOF distributions. The higher is the laser fluence, the stronger is the surface enrichment with indium and thus the larger fraction of phosphorus is evaporated in the delayed manner.

The essentially different processes responsible for indium and phosphorus vaporization during PLA of InP result in the fact that the main ablation products fly separately in the plume as seen in the experiments (Fig. 7b). Thus indium atoms and clusters have little chances to undergo collisions with $P_n$ molecules during the plume expansion while indium-indium collisions are much more probable. This explains why the clusters produced under laser ablation of InP are highly indium-rich.

The performed experiments have demonstrated that the character of laser-induced vaporization compound semiconductors (congruent or non-congruent) is much the same as that under simple thermal heating. This offers a possibility to predict the composition of the laser-generated clusters, whether they are stoichiometric or not, based on thermodynamic properties of the ablated materials. A fundamental question arises now why some of compound materials, like ZnO, vaporize congruently whereas others, e.g., InP, lose a more volatile component during vaporization. The experiments of the present work do not give an irrefragable answer to this issue which needs obviously more detailed experimental and theoretical studies of the vaporization process. We believe, however, that the difference in vaporization behavior is associated with the composition range in the phase diagram where the irradiated compound remains thermodynamically stable. The broader is the range, the higher is the loss of stoichiometry during vaporization. Physically, this probably means that such compounds with a variable composition can be structurally transformed upon heating in order to release a high-volatile component. In particular, in binary semiconductors heated up to the melting points, like-atoms can form branched chains and clusters [36]. As for the materials considered here, the ZnO crystalline structure is stable in a very narrow composition range (less than 0.1 at. % [37]) and zinc oxide evaporates congruently. In contrast, indium phosphide remains stable over a wide range of the In/P ratio (at least from 0 to 55 mole % phosphorus [38]) that results in strongly non-congruent vaporization.

## ACKNOWLEDGMENTS

It is a pleasure to thank N.M. Bulgakova and J. Bonse for useful discussions. The InP samples were kindly provided by J. Bonse. This work was supported by the Russian Foundation for Basic Research (Project 09-02-91291-CSIC).